# Switchable metamaterial reflector/absorber for different polarized electromagnetic waves

Bo Zhu, Yijun Feng [a) b)], Juming Zhao, Ci Huang, Tian Jiang [b)]

Department of Electronic Engineering, School of Electronic Science and Engineering,

Nanjing University, Nanjing, 210093 China

**Abstract**

We demonstrate a controllable electromagnetic wave reflector/absorber for different polarizations with metamaterial involving electromagnetic resonant structures coupled with diodes. Through biasing at different voltages to turn ON and OFF the diodes, we are able to switch the structure between nearly total reflection and total absorption of a particularly polarized incident wave. By arranging orthogonally orientated resonant cells, the metamaterial can react to different polarized waves by selectively biasing the corresponding diodes. Both numerical simulations and microwave measurements have verified the performance.

---

[a)] Electronic mail: yjfeng@nju.edu.cn
[b)] Also at State Key Laboratory of Millimeter Waves, Southeast University, Nanjing 210096, P.R. China.

Polarization is an important characteristic of electromagnetic (EM) waves and is often utilized in microwave and optical systems. It is always desirable to have full control of polarization in transmission, reflection or absorption of the EM waves. Many conventional methods have been employed to manipulate EM polarization, such as using grating structures, employing the Brewster effect and birefringence effect, etc. [1] Recently, metamaterials have attracted much attention due to their designable and controllable material parameters originating from artificial inclusions of EM resonant structures [2-4]. They have been employed to manipulate EM characteristics including the polarization, such as polarization manipulation through anisotropic metamaterials [5], metamaterial polarizers [6] and polarization rotators [7-8], etc.

In this letter, we propose an approach to manipulate polarization through controllable metamaterial reflector/absorber. Recently metamaterials have been designed to perfectly absorb polarized EM waves through matching the wave impedance to that of the free space and high resonant absorption [9-13]. In our design we combine two sets of metamaterial resonant structures to manipulate EM waves of orthogonal polarizations. By tuning the EM absorption through voltage biased diodes that couple the resonant structures, we are able to control continuously the reflection and absorption for a particular EM polarization, and therefore realize a switchable metamaterial reflector/absorber for different polarizations.

The proposed structure of the metamaterial reflector/absorber is based on our previous design of a polarization-independent EM wave absorber [13]. The metamaterial consists of array of electric-LC resonators (ELCs) on the top layer of a dielectric substrate and metallic film on the bottom layer (Fig. 1(a)). The ELCs can couple strongly to the electric component of the incident wave parallel to the central metallic strip of the ELC, thus supply an independent electric response. The magnetic component of the incident wave penetrates between the top and bottom layers and generates anti-parallel surface current on the ELCs and the bottom metallic film, leading to magnetic coupling and resonance. Therefore, the electric and magnetic response can be designed individually through adjusting the dimensions of ELCs and the substrate thickness to make the wave impedance of the metamaterial matched to that of the free space approximately and to introduce high EM losses from the electric and magnetic resonances, similar to the mechanism used in [9]. As a result, the transmission and reflection of the incident wave are minimized simultaneously and the absorption is maximized within a certain frequency band.

Single ELC is a polarization dependent resonant structure. To respond to both EM polarizations of the incident waves, two sets of ELCs are employed with orthogonal orientations in the unit cell, and the two ELCs with same orientation are coupled together by microwave diodes as shown in Fig. 1(a). The resonance of the ELCs can be controlled through the bias voltage applied on the diodes through vias that connect to the bias network arranged on the back of the metallic film. At resonance, opposite charges are accumulated on the two connected arches in the ELC [13]. When a zero bias voltage is applied on the diode that sets the diode OFF with a large resistance in series with a diode capacitance, it enables opposite charges to accumulate on the arches to establish strong resonance. While when a forward bias voltage is applied on the diode that sets the diode ON with a small resistance in series with an inductance, the opposite charges on the neighboring arches are cancelled out so that the resonance is destroyed. By tuning the bias voltage from zero bias to forward bias, the EM absorption of the metamaterial structure can be adjusted continuously from high level to low level.

To explore and verify the property of the proposed metamaterial structure, we have carried out both the full wave EM simulations based on finite integration technique and the free space EM reflection measurement on a fabricated sample. The structure is designed on an FR4 substrate with metallic films on both sides. In simulations, the ELC and back metallic film is modeled as copper with a thickness of 17 μm and a frequency independent conductivity of $5.8 \times 10^7$ S/m. The FR4 substrate has a permittivity of 4.4, a loss tangent of 0.02 and a thickness of 3 mm. The optimized structure is designed to work at 3.3 GHz, which includes ELC array with a period of 7.8 mm. The ELC has inner and outer radius of 3 mm and 3.6 mm respectively, with two 0.5 mm gaps, and a central metallic strip of 0.8 mm in width. The microwave diode can be modeled as a resistor in series with a capacitor at zero bias and changes to a resistor in series with an inductor under certain forward bias. The values of the lumped elements are obtained by measuring the diode at the working frequency with a RF impedance analyzer (Agilent E4991A) under different bias voltages.

The metamaterial sample (with outer dimensions of 21 × 21 cm) is fabricated based on the optimized design by integrating the metallic structures on a FR4 dielectric substrate using standard print circuit board technique as shown in Fig. 1(b). The sample is characterized through free space reflection measurement in a microwave anechoic chamber. A vector network analyzer (Agilent E8363C) and two linear polarized horn antennas are used to transmit EM waves onto the sample and receive the reflected signals. Since the proposed metamaterial is backed by a metallic sheet, the EM wave transmission is almost zero. Only the reflection coefficient $|S_{11}|$ is measured, and the EM wave absorption is determined by $A = 1 - |S_{11}|^2$. The reflection measurement is calibrated by replacing the sample with an aluminum board of the same size as perfect reflector (unit reflection).

Firstly, we explore the tunability of EM wave absorption through biasing the diodes. Fig. 2(a) shows the measured results for normal indent EM waves under different bias voltages. When the diodes are biased at zero voltage, all diodes are OFF, and the ELCs are coupled capacitively. At zero bias, $|S_{11}|$ undergoes a minimum at 3.34 GHz. A nearly perfect absorption is achieved at 3.34 GHz with a full width at half magnitude (FWHM) of about 17%. As increasing the forward bias voltage, the reflection increases gradually and the absorption drops monotonously in the absorbing band around 3.34 GHz. This is because the diode changes gradually from a capacitive element to an inductive one with small impedance when increasing the forward bias voltage. Therefore, the coupling between ELCs becomes inductive which destroys the resonance strength. The maximum reflection amplitude (nearly unity) is finally achieved over the entire frequency range with the bias voltage reaching 0.75 V. The proposed metamaterial structure works as a tunable EM reflector/absorber under different forward biasing.

To demonstrate the physical origin of the tunable EM reflector/absorber, surface current distribution on ELCs at the peak absorption frequency under normal incidence is illustrated in Fig. 2(b). Under zero bias, the ELCs are capacitively coupled by the diodes (in OFF state). The two ELCs with central strip parallel to the electric polarization react to the EM field, which results in a strong resonating surface current on these ELCs and strong absorption of this polarized EM wave. When forward biasing the diode at 0.75 V, the two ELCs become inductively coupled by the diode (in ON state). The opposite charges on the neighboring arches are cancelled out so that the resonance of individual ELC is destroyed by the coupling diode and the surface current disappears on each ELC. This polarized EM wave is almost reflected by the metamaterial structure. The other two ELCs with central strip perpendicular to the electric polarization can not react with this polarized incident wave and do not contribute to EM wave absorption with little surface current on them.

Secondly, we explore the selective reaction of the proposed metamaterial structure to different polarized EM waves. As discussed above the ELCs with different orientations react to the orthogonally polarized incident waves separately. This feature allows us to control the reflection or absorption of certain polarized incident wave through properly biasing the diodes in corresponding rows (row A or row B in Fig. 1(a)).

When the incident electric field is polarized along *x* axis, it only reacts with ELCs coupled by diodes in row A. As shown in Fig. 3(a), the reflection undergoes a minimum (nearly zero) around 3.34 GHz and nearly perfect absorption of the *x*-polarized wave is achieved with a FWHM of 17% when the diodes in row A are zero biased (OFF). However, the reflection changes to almost unity over the entire frequency range when the diodes in row A are forward biased at 0.75 V (ON). The ON or OFF state of diodes in row B has little effect on the response of the metamaterial to the *x*-polarized incident wave. On the contrary, when the incident electric field is polarized along *y* axis, the response of the metamaterial is dominated by the state of diodes in row B. As illustrated in Fig. 3(b), similar results are obtained with high reflection or total absorption when biasing the diodes in row B ON or OFF, respectively. The measurements agree with the simulations quite well except for a slight reduction of reflection under the ON state of the diodes. Both simulation and measurement results have verified that the proposed metamaterial structure can work as a controllable EM reflector/absorber for incident waves with different orthogonal polarizations through biasing the diodes ON or OFF in row A or B, respectively. For example, we can easily generate purely *x*-polarized EM wave in the reflection of arbitrarily incident waves by switching ON the diodes in row A and OFF the diodes in row B.

Finally, we explore the performance of the proposed metamaterial at oblique EM wave incidence. Both transverse electric mode (TE mode, with electric field in the plane of the structure) and transverse magnetic mode (TM mode, with magnetic field in the plane of the structure) are tested. For TE mode, the minimum reflection increases at large incident angles when the diodes are biased OFF, indicating that the absorption is slightly degraded under oblique incidence similar to the results in [13], but still has 89% absorption at the incident angle of 50° as indicated in Fig. 4(a). When the diodes are biased ON, the metamaterial structure shows nearly identical high reflectivity for oblique incident waves up to 50°. For TM mode, Fig. 4(b) shows that nearly perfect absorption or high reflection can be achieved for incident angles up to 50° with the diodes biased OFF or ON, respectively. The fluctuation of reflection coefficient at 50° incidence under the diode's ON state is probably due to the interference of direct coupling between the two horn antennas when measuring at this wide incident angle. These results indicate that the proposed metamaterial structure can work at oblique incident angles up to 50° without severe degradation of the performance.

In conclusion, we have presented a design of controllable metamaterial structure that can manipulate EM waves of different polarizations. We demonstrate that the reflection of EM wave incident upon the structure can be controlled continuously from a nearly total reflection to a total absorption by tuning the bias voltages on the diodes that couple the ELC structures in the metamaterial. By orthogonally arranging the orientation of the ELC cells, different polarized waves can be reacted with different ELCs. Therefore, we are able to switch between reflection and absorption of a particularly polarized wave through selectively biasing the diodes ON or OFF that couple the differently orientated ELCs. Due to the scalability of metamaterials, the proposed concept can be applied at other frequencies from THz to optical region. We believe such a controllable metamaterial reflector/absorber allows us more freedom to manipulate EM wave of different polarizations, and may suggest exotic applications – in particular electronically reconfigurable devices for polarization manipulation and modulation.

This work is supported by the National Basic Research Program of China (2004CB719800) and the National Nature Science Foundation of China (60990322, 60990320, 60671002, and 60801001).

**Figure captions**

Fig. 1. (Color online) (a) Schematic view of the switchable metamaterial reflector/absorber. (b) The fabricated sample. The dashed blue square indicates one unit cell with four ELCs.

Fig. 2. (Color online) (a) Measured reflection ($|S_{11}|$) under normal incidence at different forward bias voltages (indicated in the inset). (b) Surface current density distribution on ELCs at peak absorption frequency for one particular polarization.

Fig. 3. (Color online) Simulated (solid) and measured (dash) reflection ($|S_{11}|$) for normal incident waves with electric field polarized along (a) *x* axis and (b) *y* axis under different diode states.

Fig. 4. (Color online) Measured reflection ($|S_{11}|$) under oblique incidence for (a) TE mode (electric field in the plane of the absorber) and (b) TM mode (magnetic field in the plane of the absorber).

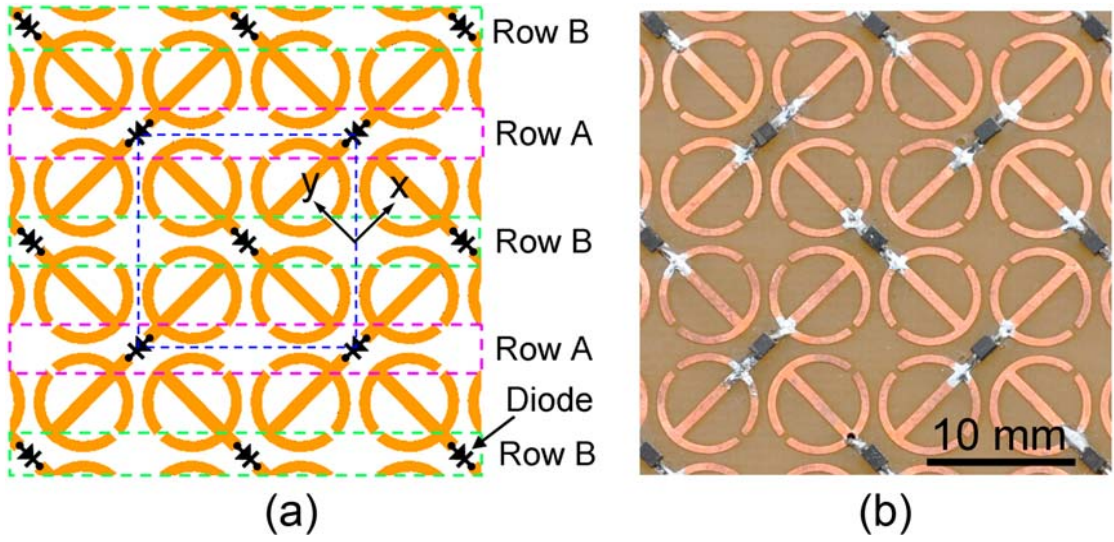

Fig. 1

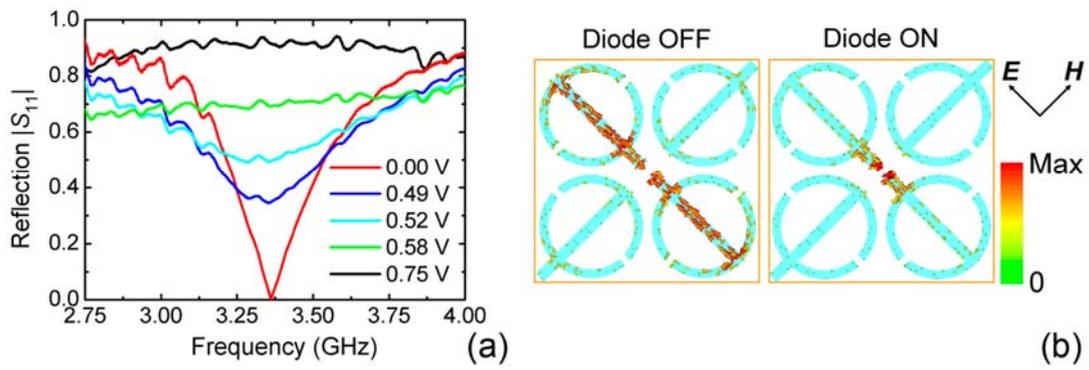

Fig. 2

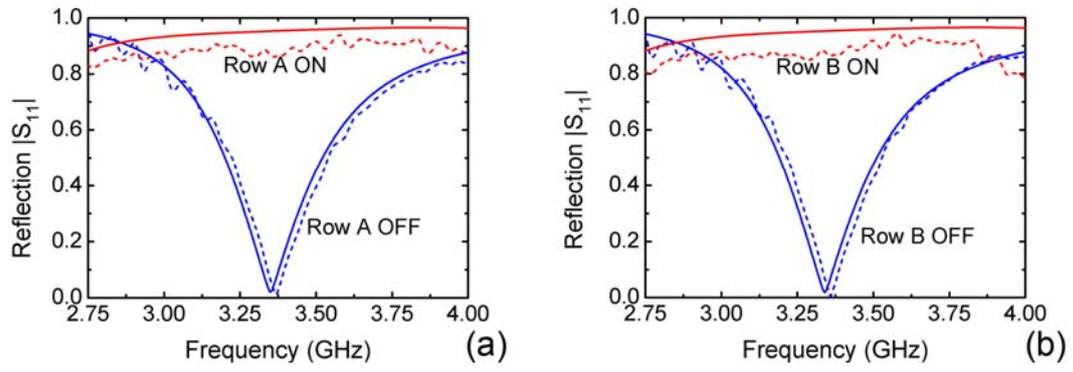

Fig. 3

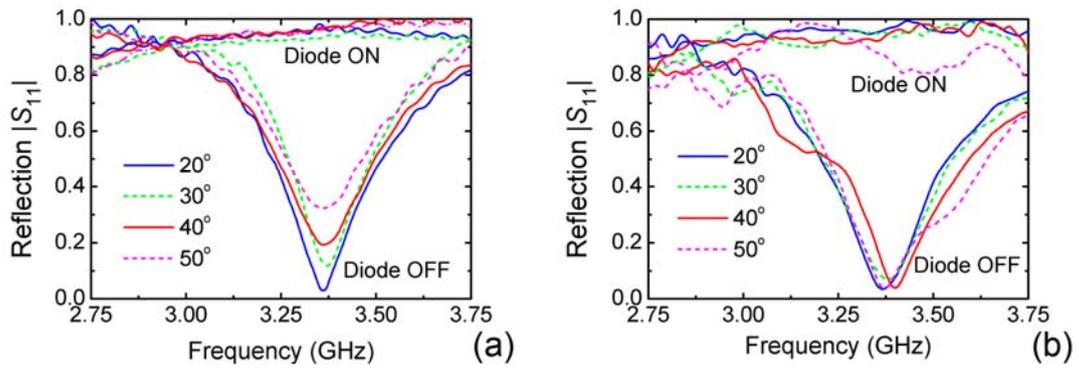

Fig. 4